\documentclass[prb,aps,twocolumn,floatfix,superscriptaddress]{revtex4-1}
\usepackage{textcomp}
\usepackage{graphicx}
\usepackage{amssymb}
\usepackage{epsfig}
\begin{document}
\title{Tuning the electronic transport properties of graphene through functionalisation with fluorine}

\author{F. Withers}
\affiliation{Centre for Graphene Science, College of Engineering, Mathematics and Physical Sciences, University of Exeter, Physics building, Exeter EX4 4QF, UK}
\author{S.Russo}
\affiliation{Centre for Graphene Science, College of Engineering, Mathematics and Physical Sciences, University of Exeter, Physics building, Exeter EX4 4QF, UK}
\author{M.Dubois}
\affiliation{Clermont Universit\'{e}, UBP, Laboratoire des Mat\'{e}riaux Inorganiques, CNRS-UMR 6002, 63177 Aubi\`{e}re, France}
\author{M.F.Craciun}
\affiliation{Centre for Graphene Science, College of Engineering, Mathematics and Physical Sciences, University of Exeter, Harrison building, Exeter EX4 4QL, UK}

\begin{abstract}

Engineering the electronic properties of
graphene has triggered great interest for potential applications in electronics and opto-electronics. Here we demonstrate the possibility to tune the electronic transport properties of graphene monolayers and multilayers by functionalisation with fluorine. We show that by adjusting the fluorine content different electronic transport regimes can be accessed. For monolayer samples, with increasing the fluorine content, we observe a transition from electronic transport through Mott variable range hopping in two dimensions to Efros - Shklovskii variable range hopping. Multilayer fluorinated graphene with high concentration of fluorine show two-dimensional Mott variable range hopping transport, whereas CF$_{0.28}$ multilayer flakes have a band gap of 0.25eV and exhibit thermally activated transport. Our experimental findings demonstrate that the ability to control the degree of functionalisation of graphene is instrumental to engineer different electronic properties in graphene materials.

\end{abstract}

\maketitle
Graphene, a monolayer of sp2 bonded carbon atoms arranged in a honeycomb pattern, is a two-dimensional semimetal where the valence and conduction bands touch in two independent points at the border of the Brillouin zone, named K and K' valleys \cite{Wallace,CastroNeto,Novoselov2004,Novoselov2005,Zhang2005}. This material has remarkable electronic, optical and mechanical properties which can be used in a new generation of devices \cite{Geim-Novoselov2007,Geim2009}. For instance, the high mobility of charge carriers, is attracting considerable interest in the realm of high-speed electronics \cite{Morozov2008}. Furthermore, thanks to the unique combination of high electrical conductivity \cite{Novoselov2005,Zhang2005} and optical transparency \cite{Nair2008}, graphene is a promising material for optoelectronic applications such as displays, photovoltaic cells and light-emitting diodes. Few-layer graphene are yet unique materials \cite{Craciun2011} with unprecedented physical properties: bilayers are semiconductors with a gate-tuneable band gap \cite{Ohta2006,Castro2007,Oostinga2008,LMZhang2008,Zhou2009,Zhang2009,Mak2009,Kuzmenko2009,Russo2009,Xia2010,Zou2010}, whereas trilayers are semimetals with a gate-tuneable overlap between the conduction and valence bands \cite{Craciun2009,Koshino2009}. However, the use of graphene for applications in daily-life electronics suffers from a major drawback, that is the current in graphene cannot be simply pinched off by means of a gate voltage. A valuable solution to this problem is to engineer a band gap in the energy spectrum of graphene for example confining the physical dimensions of graphene into nanoribbons \cite{Son2006,Li2008,Jiao2009,Oostinga2010} or by chemical functionalisation \cite{Sofo2007,Boukhvalov2009,Leenaerts2010,Sahin2011,Elias2009,Ryu2008,Balog2010,Worsley2007,BittoloBon2009,Withers2010,Nair2010,Robinson2010,Cheng2010,Jeon2011,Hong2011,Dikin2007,Park2009,Eda2010}.

When chemical elements -e.g. oxygen, hydrogen, or fluorine- are adsorbed on the surface of graphene they form covalent bonds with the carbon atoms. As a result, the planar crystal structure of graphene characterized by sp2 bonds between the carbon atoms is transformed into a three-dimensional structure with sp3 bonds. The adsorbed elements can attach to graphene in a random way, as it is the case in graphene oxide \cite{Dikin2007,Park2009,Eda2010}, or they can form ordered patterns as it has been found for hydrogen \cite{Elias2009,Ryu2008,Balog2010} and fluorine \cite{Worsley2007,BittoloBon2009,Withers2010,Nair2010,Robinson2010,Cheng2010,Jeon2011,Hong2011} adsorbates. \textit{Ab initio} calculations performed within the density functional theory formalism predict that functionalisation with hydrogen and fluorine lead to an expected band gap of 3.8 eV and respectively 4.2 eV for full functionalisation \cite{Sofo2007,Boukhvalov2009,Leenaerts2010,Sahin2011}.

Successful hydrogenation and fluorination of graphene has been recently achieved by several groups \cite{Elias2009,Ryu2008,Balog2010,Worsley2007,BittoloBon2009,Withers2010,Nair2010,Robinson2010,Cheng2010,Jeon2011,Hong2011}. Hydrogenation is usually carried out in a remote plasma of $H_{2}$ \cite{Elias2009,Ryu2008,Balog2010} which makes it difficult to control the degree of induced atomic defects as well as the stoichiometry of the functionalisation. Furthermore, hydrogenated graphene can loose H at moderate temperatures \cite{Elias2009}, which limits the use of this material in applications where high temperature stability is required. On the other hand, fluorine has higher binding energy to carbon and higher desorption energy than hydrogen \cite{Sofo2007,Boukhvalov2009,Leenaerts2010,Sahin2011}. Opposed to hydrogenation, the process of fluorination is easy to control -e.g. \textsl{via} temperature and reactant gases- leading to reproducibly precise C/F stoichiometries.

Here we explore the electronic properties of graphene functionalised with a fluorine content ranging from 7\% to 100\%. We have fabricated transistor structures with fluorinated graphene monolayers and multilayers and studied their electrical transport properties in the temperature range from 4.2K to 300K. We show that the electronic transport properties of fluorinated graphene can be tuned by adjusting the fluorine content, so that different transport regimes can be accessed, i.e. Mott variable range hopping in two dimensions \cite{Mott1969,Shklovskii1984}, Efros-Shklovskii variable range hopping \cite{Efros1985} and thermally activated transport.

\begin{figure}
\resizebox{0.45\textwidth}{!}{%
  \includegraphics{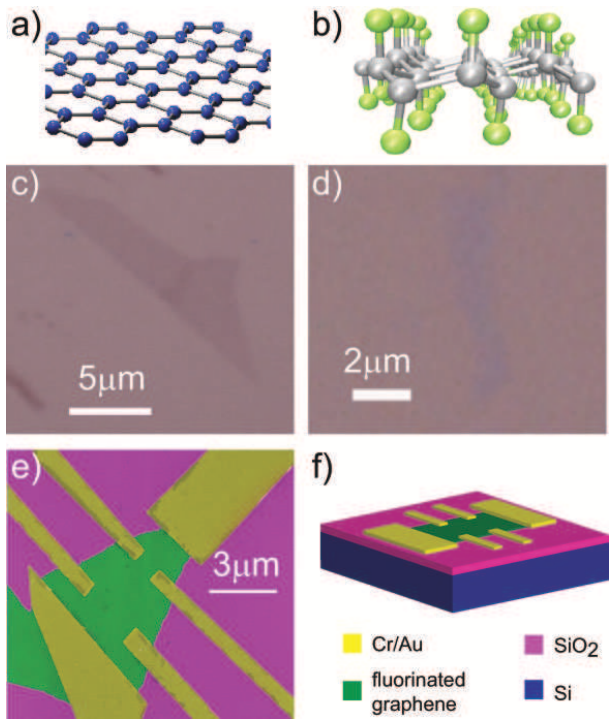}
}
\caption{Crystal structure of pristine graphene (a) and fluorinated graphene (b). The grey balls in (b) represent the carbon atoms, whereas the green balls are the fluorine atoms. Optical image of pristine graphene (c) and of fluorinated graphene (d).
(e) False color SEM image of a fluorinated graphene device. (f) Schematic view of the transistor structure fabricated on fluorinated graphene.}
\label{fig:1}       
\end{figure}

To produce fluorinated graphite we have used two distinct methods \cite{Nakajima1995,Touhara2000,Sato2004,Zhang2010}. In the first method graphite is heated in the presence of F$_2$ to temperatures in excess of 300\textcelsius, so that covalent C-F bonds are formed and modify the carbon hybridisation \cite{Nakajima1995}. The layered structure of graphite is then transformed into a three-dimensional arrangement of carbon atoms. In this work we present studies on graphene exfoliated from fully fluorinated HOPG graphite CF$_{n}$ and CF$_{0.28}$ synthesized with this method. However due to the harsh fluorination conditions, many structural defects are formed, which makes it very difficult to exfoliate large enough monolayer flakes that can be identified by optical microscopy and easily processed into devices. To prepare larger fluorinated graphene samples, we have used a second fluorination method where graphite is exposed to a fluorinating agent -i.e. XeF$_{2}$. In this case the functionalisation process is carried out at $T \leq 120$\textcelsius, as XeF$_2$ easily decomposes on the graphite surface into atomic fluorine \cite{Sato2004}. The mixture of natural graphite and XeF$_2$ was prepared in a glove box in an Ar atmosphere. Due to its reactivity and diffusion, the fluorination results in a homogenous dispersion of fluorine atoms that become covalently bonded to carbon atoms \cite{Sato2004,Zhang2010,Zhang2008}. At low fluorine content, the F/C atomic ratio is $\leq$ 0.4. In this case the conjugated C-C double bonds in the non-fluorinated parts and covalent C-F bonds in corrugated fluorocarbon regions coexist \cite{Zhang2010,Giraudet2007}. The concentration of the covalent bonds increases with increasing the concentration of fluorine. The samples produced using the XeF$_2$ gas that we investigate here have the concentration of fluorine of 7\%, 24\% and 28\%.

We fabricated transistor structures by mechanical exfoliation of the fluorinated flakes onto conventional Si/SiO2(275nm) substrates. Flakes are located using an optical microscope (see Figure 1 (d)) and subsequently characterized by Raman spectroscopy. Monolayer graphene flakes were identified by fitting the 2D peak of the Raman spectra by a single Lorentzian function (see Fig. 2b), with a full width at half maximum (FWHM) of 30-45 cm$^{-1}$ which is typical for pristine monolayer graphene \cite{Ferrari2006}. In total, four monolayer flakes were processed into four-terminal transistor devices where the electrical contacts were defined by e-beam lithography, deposition of
Cr/Au (5 nm/50 nm) and lift-off procedure, see Figure 1c.

The typical optical contrast of fluorinated graphene is $\sim 2-6\%$, which is systematically lower
than what we observe on pristine graphene ($\sim 9 \%$). The reduced contrast in fluorinated graphene has to be expected, since the opening of a large energy-gap in the energy dispersion of fluorinated graphene lowers the optical absorption transitions between conduction and valence bands.

\begin{figure}
\resizebox{0.5\textwidth}{!}{%
  \includegraphics{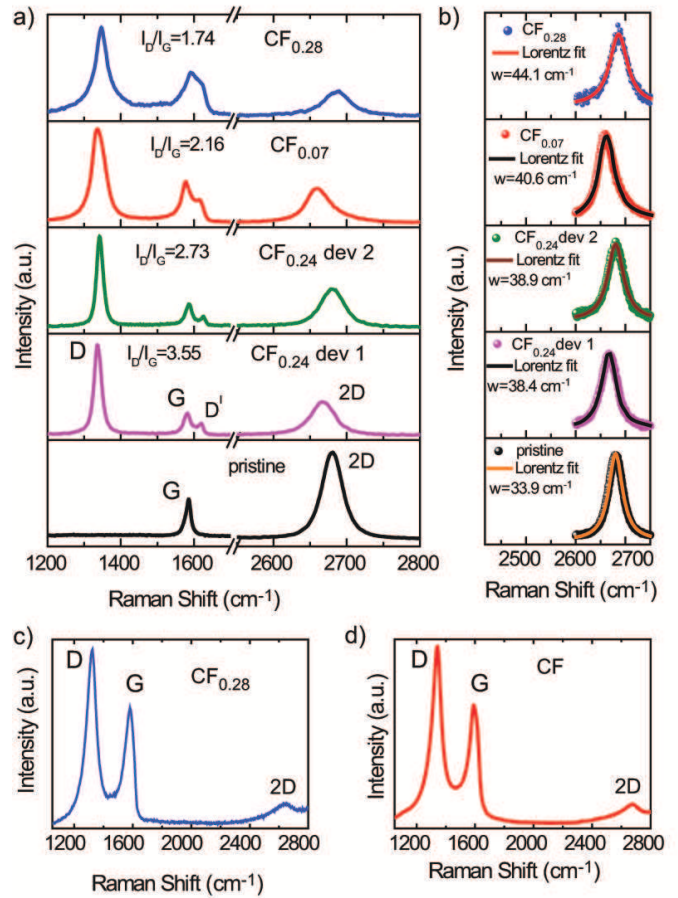}
}
\caption{(a) Raman spectra of monolayer fluorinated graphene and pristine monolayer graphene. (b)
Fitting of the 2D peak with a single Lorentzian function for pristine and fluorinated monolayer graphene.
(c) and (d) Raman spectra of fluorinated multilayer graphene.}
\label{fig:2}       
\end{figure}

We have characterized all the exfoliated flakes by Raman spectroscopy using an excitation light with a wavelength of 532 nm and a spot size of 1.5 $\mu$m in diameter. An incident power of 5 mW was used. We ensured that this power does not damage the graphene by performing Raman measurements on a similarly sized pristine graphene flake which shows the common spectra of mechanically exfoliated graphene: the G band and 2D band (also known as G') at 1580 cm$^{-1}$ and 2700 cm$^{-1}$, see Figure 2. The G band is associated with the double degenerate E2g phonon mode at the Brillouin Zone center, while the 2D mode originates from a second order process, involving two intervalley phonons near the K point, without the presence of any kind of disorder or defect \cite{Ferrari2006}. In the fluorinated graphene, additional peaks are activated in the Raman spectra (see Figure 2), the D and D' peaks that appear at 1350 cm$^{-1}$ and 1620 cm$^{-1}$. These Raman peaks originate from double-resonance processes at the K point in the presence of defects, involving respectively intervalley (D) and intravalley (D') phonons \cite{Dresselhaus2010,Pimenta2007,MartinsFerreira2010,Lucchese2010}.

In exfoliated pristine graphene, the D peak can only be observed at the edges of the flakes where there is a large concentration of structural defects and its intensity is typically much lower than the intensity of the G peak \cite{Casiraghi2009,Malard2009}. In our studies performed on pristine graphene flakes with similar size as the fluorinated graphene flakes, the intensity of the D-peak is typically well below the sensitivity of our Raman setup -i.e. we are usually not able to detect any D-peak due to the edges of the flakes. Therefore, the observed D-peak in our fluorinated graphene samples must originate from other defects than simply the edges of the samples. As all our samples contain networks of sp2 bonded carbon atom rings, we believe that the D peak is mainly activated by the F atoms which act as vacancies in these sp2 rings.

A better understanding of the level of disorder in our samples is reached when analysing the intensity ratio I$_{D}$/I$_{G}$ for the D band and G band. It has been recently shown that in graphene I$_{D}$/I$_{G}$ has a non-monotonic dependence on the average distance between defects L$_{D}$, increasing with increasing L$_{D}$ up to L$_{D}$ $\sim$ 4nm and decreasing for L$_{D}$ $>$ 4nm \cite{MartinsFerreira2010,Lucchese2010,Cancado}. Such behavior has been explained by the existence of two disorder-induced regions contributing to the D peak: a structurally disordered region of a radius $\sim$ 1nm around the defect and a larger defect-activated region which extends to $\sim$ 3nm around the defect. In the defect activated region, the lattice structure is preserved, but the proximity to a defect causes a mixing of Bloch states near the K and K' valleys. Consequently, the breaking of the selection rules leads to an enhancement of the D peak. Furthermore, it was shown that in the structurally disordered region, the G and D' peaks overlap.

The Raman spectra of fluorinated monolayer samples produced from graphite with fluorine content of 7\% and 28\% (see Fig. 2a) systematically shows that the G and D' peaks have a significant overlap. On the other hand, the samples exfoliated from CF$_{0.24}$ exhibit very distinct G and D' peaks. Based on the aforementioned phenomenological model \cite{Lucchese2010} we can state that the CF$_{0.07}$ and CF$_{0.28}$ samples are in the regime where the intensity ratio I$_{D}$/I$_{G}$ increases with increasing L$_{D}$ (i.e. decreasing the concentration of F) whereas the CF$_{0.24}$ samples are in the opposite regime. This scenario is confirmed when comparing the intensity ratio I$_{D}$/I$_{G}$ for fluorinated monolayers extracted from graphite with different fluorine content: I$_{D}$/I$_{G}$=1.74 for CF$_{0.28}$ and I$_{D}$/I$_{G}$= 2.16 for CF$_{0.07}$. Furthermore, from the L$_{D}$ dependence on I$_{D}$/I$_{G}$ we estimate L$_{D}$ $\sim$ 1.5 nm for CF$_{0.28}$ and L$_{D}$ $\sim$ 2 nm for CF$_{0.07}$ \cite{Lucchese2010}. For the samples exfoliated from CF$_{0.24}$ we estimate L$_{D}$ $\sim$ 5.3 nm for CF$_{0.24}$ (device 1) and L$_{D}$ $\sim$ 6.1 nm for CF$_{0.24}$ (device 2). These values for L$_{D}$ are in agreement with the observed frequencies of the D, G, D' and 2D peaks as well as with the FWHM values of the 2D peaks \cite{MartinsFerreira2010}.

In the case of the fluorinated multilayer flakes, see Figure 2 (c) and (d), it is difficult to perform a similar analysis, as the intensity of the G band depends on the number of graphene layers present in the sample \cite{Ferrari2006}. For samples thicker than 3-4 layers, the structure of the 2D peak does not provide an accurate estimation for the number of layers due to the large number of fitting parameters.

\begin{figure}
\resizebox{0.5\textwidth}{!}{%
  \includegraphics{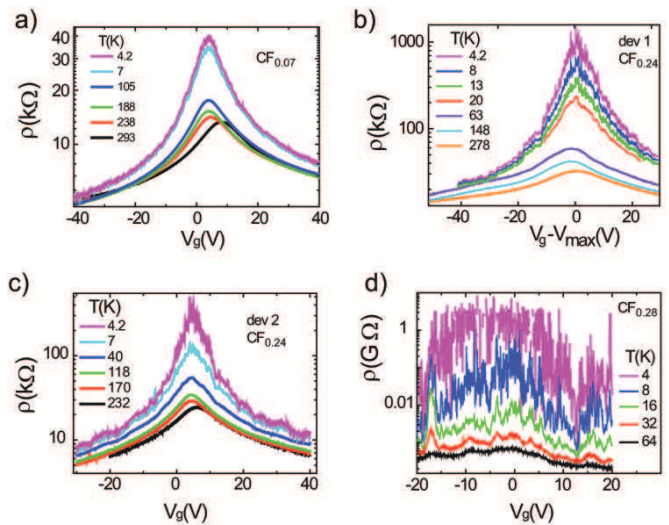}
}
\caption{Resistivity of 4 fluorinated monolayer graphene samples as a function of gate voltage at different temperatures.
The different panels correspond to different concentration of fluorine in the starting fluorinated graphite material,
as indicated in each panel.}
\label{fig:3}       
\end{figure}

Having characterized the level of disorder from Raman spectroscopy, we now proceed to address the role of disorder on the electrical transport properties of fluorinated graphene materials. The resistance of the transistor devices was measured both in dc, by means of Keithley 2400 Source-meter, and in ac at low frequency (34Hz) with a lock-in amplifier in a voltage-biased configuration. For the ac-measurements, the excitation current was varied to
ensure that the resulting voltage was smaller than the temperature
to prevent heating of the electrons and the occurrence of
nonequilibrium effects. The comparison of 2- and 4-probe
measurements shows that the contact resistance in our devices
is negligible as compared to the sample resistance. This experimental finding insures that even 2-probe transport measurements are probing the electrical properties of the bulk fluorinated graphene rather then simply the Cr/graphene interface.

Figure 3 shows the resistivity ($\rho$) as a function of gate voltage (V$_{g}$) for the fluorinated monolayer samples for different temperatures. The resistivity exhibits a non monotonous dependence on $V_{g}$ with a maximum at V$_{g}$ =+10 V, stemming for a doping level of $n = 0.74 \cdotp 10^{12} cm^{-2}$ commonly seen also in pristine graphene devices and attributed to doping by atmospheric water. In all cases the resistivity of fluorinated graphene shows a pronounced temperature dependence. Indeed, the maximum of resistivity changes over two orders of magnitude as T decreases from 300 K to 4.2 K. Away from the maximum of resistivity region the temperature dependence remains weak, with the mobility of carriers of 150 cm$^{2}$/Vs. Furthermore, at low temperature the resistance shows strong mesoscopic fluctuations, as expected for samples of small size \cite{Kechedzhi2009}. In the analysis of the maximum of resistivity we smooth the $\rho$(V$_{g}$) curves using a moving average filter.

\begin{figure}
\resizebox{0.5\textwidth}{!}{%
  \includegraphics{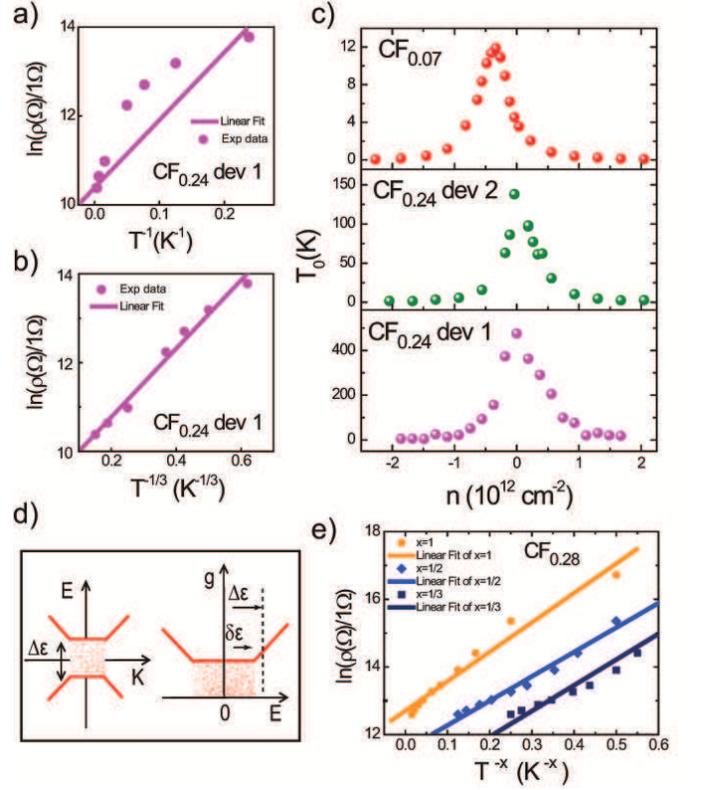}
}
\caption{Resistivity of fluorinated monolayer graphene in the charge neutrality region plotted as a function of T$^{-1}$ (a) and T$^{-1/3}$ (b). (c) The values of the hopping parameter T$_{0}$ as a function of carrier density for the samples where transport occurs by two-dimensional Mott variable range hopping.
(d) Schematic diagrams of the energy dispersion of fluorinated monolayer graphene (left panel) and of the energy dependence of the density of electron states, with the Fermi level at zero energy (right panel). The localised states are shown by the shaded area. (e) Resistivity of fluorinated monolayer graphene (exfoliated from CF$_{0.28}$ graphite)
in the charge neutrality region plotted as a function of T$^{-x}$. The solid lines represent fits to the experimental
data where x=1 for thermally activated transport, x=1/2 for Efros-Shklovskii variable range hopping and x=1/3 for two-dimensional Mott variable range hopping. The best fit is obtained for x=1/2.
}
\label{fig:4}       
\end{figure}

To examine the presence of the energy gap we analyse the temperature dependence of the maximum of resistivity by an exponential law describing thermal activation of carriers across an energy gap $\Delta\varepsilon$:
$\rho(T)=\rho_{0}exp(\Delta\varepsilon/2k_{B}T$), see Figure 4(a). This analysis clearly shows that our data are not described by the thermal activation law over the whole temperature range. We note that the slope of $ln\rho(1/T)$ \textsl{versus} $1/T$ decreases with decreasing T, which is a signature of hopping conduction via localised states \cite{Shklovskii1984}. The fact that in the whole range of studied temperatures electron transport is not due to thermal activation across the gap but due to hopping becomes clear when re-analysing the temperature dependence in terms of the two-dimensional Mott variable range hopping (2D-VRH) \cite{Mott1969,Shklovskii1984}. In this model the functional dependence of $\rho$ on temperature is
$\rho(T)=\rho_{0}exp(T_{0}/T)^{1/3}$, where $k_{B}T_{0} = 13.6/a^{2}g(\mu)$, g is
the density of localised states at the Fermi level $\mu$ and a is the localisation length \cite{Mott1969,Shklovskii1984}. Experimentally we find that the measured $\rho(T)$ for the samples produced from CF$_{0.07}$ and CF$_{0.24}$ graphite (see Figure 3(b) for CF$_{0.24}$) is described well by the 2D-VRH model.

Figure 4(c) shows the hopping parameter T$_{0}$ as a function of carrier concentration for these 3 samples. The value of T$_{0}$ approaches zero at a carrier concentration of $\pm 1.2 \cdotp 10^{12} cm^{-2}$. This value gives the concentration of the localised electron states in the energy range from $\varepsilon = 0$ to the mobility edge, see Figure 4(d). The mobility edge occurs at $V_{g} \pm 20 V$ and indicates the transition from hopping to metallic conduction.

In order to relate the obtained concentration of the localised states to the energy gap $\Delta\varepsilon$, one needs to know the exact energy dependence of the density of states in the gap. For estimations, we will use the linear relation for the density of extended states above the mobility edge $g(\varepsilon) = 2 \varepsilon/\pi\hbar^{2}v^{2}$ ($v = 10^{6} m/s$ is the Fermi velocity), and a constant value for the density of localised states below the mobility edge, Figure 4(d). This gives an estimation $\Delta\varepsilon$=60 meV and twice this value for the full mobility gap. In this approximation the density of the localised states in the gap is estimated as $10^{36} J^{-1}m^{-2}$. Using the obtained value of the hopping parameter at the maximum of resistivity, we can then estimate the localisation length at $\varepsilon = 0$ as a=40 nm for CF$_{0.24}$ (device 1), a=81 nm for CF$_{0.24}$ (device 2) and a=265nm for CF$_{0.07}$.

Figure 4(e) shows the analysis of the temperature dependence of the resistivity for fluorinated monolayer graphene exfoliated from CF$_{0.28}$ graphite. For this sample, characterised by the largest disorder $L_{D}\sim 1.5nm$, the experimental data cannot be described by thermally activated law nor Mott variable range hopping. In this case, the $ln(\rho)$ follows the typical $T^{-1/2}$ dependence described by Efros-Shklovskii variable range hopping in the presence of Coulomb interaction between the localised states ($\rho(T)=\rho_{0}exp(T_{0}/T)^{1/2}$) \cite{Efros1985}.
T$_{0}$ is related to the localisation lengths through T$_{0}=2.8e^{2}/4\pi\varepsilon\varepsilon_{0}k_{B}a$ and for our sample we estimate T$_{0}=52K$. Assuming that $\varepsilon$ is the dielectric constant of SiO$_{2}$ we obtain the localisation length a=282 nm.

We turn now our discussion to multilayer fluorinated graphene exfoliated from fully fluorinated graphite and from CF$_{0.28}$ prepared by exposure to fluorine gas. The fully fluorinated multilayer show systematically a very large resistance (more than 100 GOhm) and no gate-voltage control of the resistivity. To achieve gate modulation in these samples, we reduced the fluorine content by annealing the samples at 300\textcelsius in a 10 $\%$ atmosphere of H$_{2}$/Ar for 2 hrs. After this processing the resistance is decreased and a partial gate-voltage control is achieved, Figure 5(a). The annealing, however, has not noticeably changed the Raman spectrum, see Figure 2(d).

Resistance measurements of CF$_{n}$ flakes after annealing show a strong temperature dependence, Figure 5(a). Analysis of the temperature dependence of the resistance in terms of the activation law at the highest gate voltage V$_{g}$ =50 V, (which is still far from the Dirac point) gives a gap of only 25 meV, which is significantly smaller than the expected energy gap for fully fluorinated graphene. Similarly to the fluorinated monolayer graphene, the resistivity dependence on temperature is fitted well by variable range hopping with the value of $T_{0} =20000 K$.This confirms that the previously found activation energy of 25 meV is not the activation energy $\Delta\varepsilon$ that separates the localised states from extended states at the mobility edge, but is an activation energy $\delta\varepsilon$ of hopping between localised states within the mobility gap, see Figure 4(d).

Figure 5(c) and (d) show the transport data for the multilayer fluorinated graphene CF$_{0.28}$ prepared by exposure to fluorine gas. The I-V characteristics of these samples are strongly non-linear (see Figure 5c) with resistances of more than 1 $G\Omega$. This truly insulating state is compatible with an energy band gap in the fluorinated graphene. Furthermore, the presence of a band gap is demonstrated by the dependence of the resistivity on temperature which is described well by a thermally activated law with a $\Delta\varepsilon$=0.25eV (see Figure 5d).

\begin{figure}
\resizebox{0.5\textwidth}{!}{%
  \includegraphics{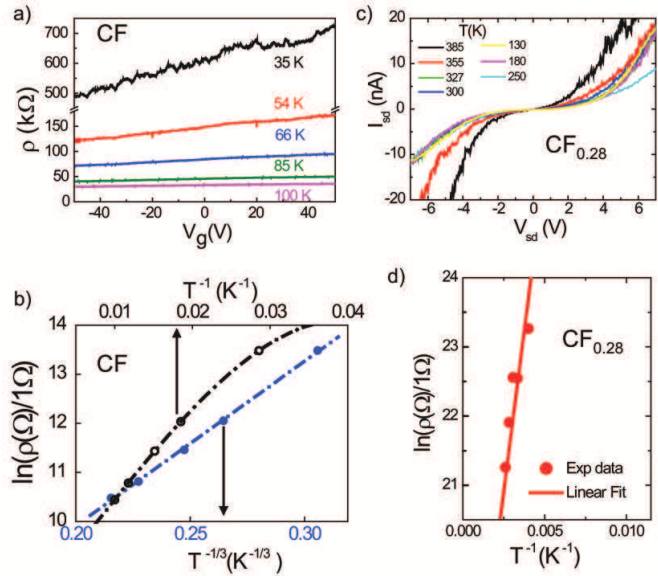}
}
\caption{Transport in fluorinated multilayer samples. (a) Resistivity as a function of gate
voltage for an annealed fully fluorinated device. (b) Resistivity for the fully fluorinated device plotted against T$^{-1}$ and T$^{-1/3}$ at V$_{g}$=+50 V. (c) I-V characteristics for multilayer fluorinated graphene exfoliated from CF$_{0.28}$ graphite.
(d) Resistivity of fluorinated multilayer graphene (exfoliated from CF$_{0.28}$) plotted as a function of T$^{-1}$}
\label{fig:5}       
\end{figure}

In conclusion, we have demonstrated the possibility to tune the band structure and therefore the electronic transport properties of graphene through functionalisation with fluorine. In particular, depending on the fluorine concentration different transport regimes can be accessed. For monolayer samples we observe a transition from two-dimensional Mott variable range hopping to Efros-Shklovskii variable range hopping with increasing the fluorine content. Multilayer fluorinated graphene with high concentration of fluorine show two-dimensional Mott variable range hopping, whereas CF$_{0.28}$ multilayer flakes have a gap of 0.25eV and exhibit thermally activated transport. Our experimental findings demonstrate that the ability to control the degree of functionalisation of graphene is instrumental to engineer different electronic properties in graphene materials.  In all cases fluorinated graphene transistors exhibit a large on/off ratio of the current, making this material of interest for future applications in transparent and bendable electronics.

S.R. and M.F.C. acknowledge financial support from EPSRC (Grant no. EP/G036101/1 and no. EP/J000396/1). S.R. acknowledges financial support from the Royal Society Research Grant 2010/R2 (Grant no. SH-05052).

\end{document}